# Realization of ppm level pressure stability for primary thermometry using a primary piston gauge

Bo Gao[1,*,&], Hui Chen[2,1,&], Dongxu Han[3,&], Pascal Gambette[4,&], Haiyang Zhang[1], Changzhao Pan[4], Yingwen Liu[2], Bo Yu[3], Ercang Luo[1], Mark Plimmer[4], Laurent Pitre[4]

1, Technical Institute of Physics and Chemistry (TIPC), Chinese Academy of Sciences (CAS), Beijing 100190, China

2, Key Laboratory of Thermo-Fluid Science and Engineering of MOE, School of Energy and Power Engineering, Xi'an Jiaotong University, Xi'an 710049, China

3, School of Mechanical Engineering, Beijing Institute of Petrochemical Technology, Beijing 102617, China

4, Laboratoire national de métrologie et d'essais-Conservatoire national des arts et métiers (LNE-Cnam), La Plaine-Saint Denis F93210, France

*Corresponding author, Tel: +86-10-82543730,

E-mail: bgao@mail.ipc.ac.cn

[&]These authors contributed to the work equally and should be regarded as co-first authors.

## Abstract

To achieve an uncertainty of 0.25 mK in single-pressure refractive-index gas thermometry (SPRIGT), the relative pressure variation of He-4 gas in the range 30 kPa to 90 kPa, should not exceed 4 ppm ($k$=1). To this end, a novel pressure control system has been developed. It consists of two main parts: a piston gauge to control the pressure, and a home-made gas compensation system to supplement the micro-leak of the piston gauge. In addition, to maintain the piston at constant height, a servo loop is used that automatically determines in real time the amount of extra gas required. At room temperature, the standard deviations of the stabilized pressure are 3.0 mPa at 30 kPa, 4.5 mPa at 60 kPa and 2 mPa at 90 kPa. For the temperature region 5 K-25 K used for SPRIGT in the present work, the relative pressure stability is better than 0.16 ppm i.e. 25 times better than required. Moreover, the same pressure stabilization system is readily transposable to other primary gas thermometers.

## 1 Introduction

The 1990 international practical temperature scale (ITS-90, $T_{90}$) [1] is used worldwide to approximate thermodynamic temperature $T$. The scale is based on the best thermodynamic temperature measurements that were available in 1990. To measure the temperature in the range 5 K to 24.5561 K (the neon triple point), three primary thermometry techniques are traditionally employed: constant volume gas thermometry (CVGT) [2,3], dielectric constant gas thermometry (DCGT) [2] and acoustic gas thermometry (AGT) [2,4,5]. The core principle of CVGT is to measure the absolute pressure in a constant volume. In DCGT, the thermodynamic temperature is obtained from measurements of the dielectric constant of a gas. In AGT, it is the speed of sound in a monatomic gas that yields thermodynamic temperature.

SPRIGT, by contrast, is as a relative primary gas thermometry method, whereby the thermodynamic temperature is determined by microwave frequency measurements in a gas-filled resonator. The resonance frequency is measured over the temperature range of interest first with the resonator under vacuum (in practice $p$ =10 μPa). The thermometric gas is then admitted and its pressure maintained constant at the chosen value. Microwave resonant frequencies are measured with the gas present at different temperatures[6].

Taking advantage of recent advances in the *ab initio* calculation of the thermophysical properties of He-4, SPRIGT has the potential to yield an uncertainty as small as 0.25 mK. A suitable cryostat for it was developed in our group and details of the temperature control system, thermal response characteristics of the cryostat and microwave measurement published [7-10]. However, stable long-term pressure is also required i.e. one constant to better than 0.12 Pa over periods of days or weeks, a considerable challenge for any pressure control system. The design, development and performance of such a system at both ambient and cryogenic temperatures are the subject of the present article, only a short report on the pressure control at room temperature only having been published up to now[11].

The pressure stabilization scheme employed in our work exploits a piston gauge in two novel ways. Piston gauges (PGs) are traditionally used as primary and secondary pressure standards in national metrology institutes [12-17]. To achieve long-term pressure stability, many researchers first calibrate a pressure transducer using a piston pressure balance, then use the calibrated gauge output as the input for a pressure control feedback loop [18-21]. By contrast, in the system described here, the piston gauge is used directly for pressure control, thereby omitting the intermediate steps of pressure value transfer, which leads to more accurate pressure measurement and closed-loop stability.

The second innovation concerns the mode of operation of the piston gauge. In the traditional working mode, the loaded piston rotates freely inside the cylinder, behaving as an aerodynamic bearing. Sutton and Fitzgerald have discussed the aspects of rotated gas-operated pressure balances that affect their performance [22]. It is recommended that the piston gauge be operated with a small, reproducible falling rate [14]. Izumi *et al.* [23] evaluated the falling rate of a liquid-lubricated piston–cylinder (PC) assembly. In the work of Sabuga *et al.* [24], the gap width of the PC assembly at PTB was adjusted to yield an optimal compromise between the rate of descent and sensitivity.

However, when the piston gauge employed in the present work (Fluke PG 7601) was used in rotating mode, unwanted peaks appeared in the pressure profile due to the acceleration of the rotating piston whenever its speed reached its lower limit. Moreover, each time the piston had fallen for one or two hours, it would reach its lowest permissible height and so no longer be able to control the pressure. (To add pressure to the system, the piston must drop). To prevent this from being a problem, therefore, we should not only isolate the pressure oscillation from the surroundings (external pressure-noise), but more importantly isolate the pressure change due to the piston gauge (internal pressure-noise).

Bock *et al.* at PTB [25] discovered that by using a large area *non-rotating* piston gauge, they were able to reduce the uncertainty of their vacuum pressure scale. This approach was followed in the present work. In addition, a novel piston suspension gas path compensation system was implemented to prevent the piston from reaching the bottom of the cylinder.

This article presents experimental results for pressure control using this system. To gain a better understanding of the pressure control principle, the pressure control was first implemented using the internal height sensor of the commercial piston gauge. For finer height measurement and control, an external laser interferometer was used to determine the height of the upper surface and enabled ultra-stable pressure control in the range of 30-90 kPa in a ballast volume at room temperature. Finally, results are given for different pressures (30 kPa, 60 kPa and 90 kPa) and low temperatures (5 K-25 K) when the pressure control system was connected to the resonator in the cryostat.

## 2 Apparatus

An overview of the apparatus is given in Fig. 1 (a). The three main components are the temperature control box, the piston gauge (Fig. 1 (b)) and the gas compensation system (ballast volume, mass flow controller, absolute pressure gauge).

## 2.1 Temperature control box

The temperature control box was designed to provide a stable working temperature for the instruments (24.5°C, temperature instability <1.5 mK) to reduce the impact of external temperature variation on the measured pressure, the temperature box is based on the description given in Berg *et al* [26].

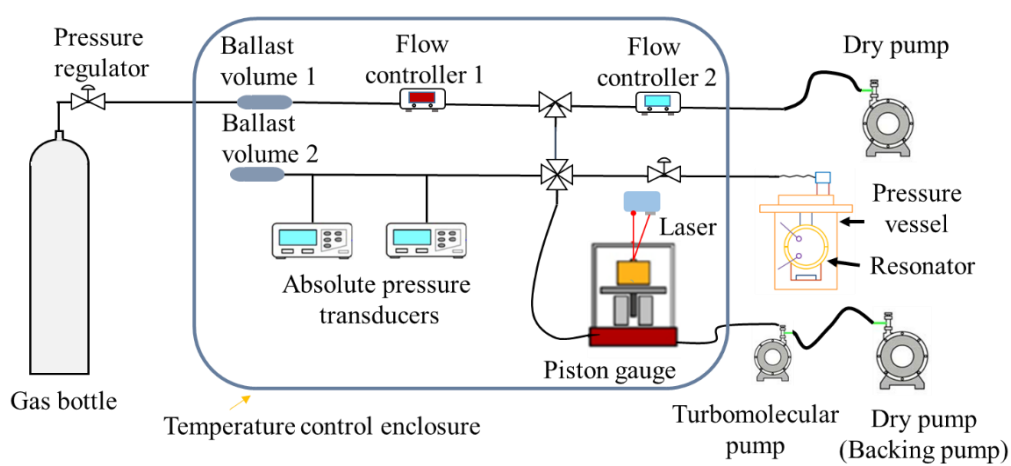

(a)

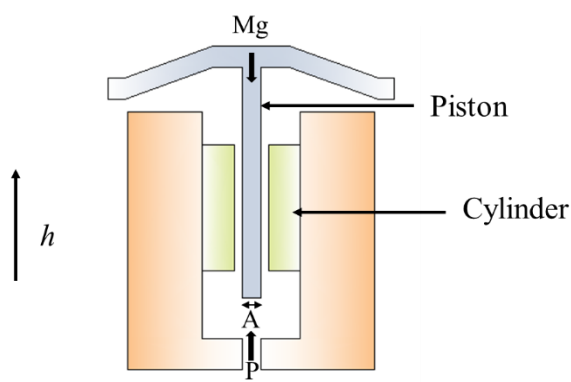

(b)

**Fig. 1 The schematic diagram of the pressure control system (a) Gas compensation system (b) Gas-operated piston gauge**

## 2.2 Piston gauge

Piston gauges are oft-used standard instruments in pressure metrology [27]. Fig. 1

(b) shows a schematic of a gas-operated piston gauge, which consists of a loaded piston in a close-fitting vertical cylinder. The piston moves vertically in the cylinder, changing the pressure in the volume beneath the piston until the gravitational force on the piston and its load are balanced by the force of pressure acting on the lower surface of the piston.

The absolute pressure $P_A$ beneath the piston is given by the force balance equation

$$P_A = \frac{Mg}{A_{\theta,p}} + P_{HA} + P_{vac} \tag{1}$$

where $A_{\theta,P}$ the effective area of the piston-cylinder, $M$ the mass load, $P_{HA}$ the fluid head correction, $g$ the local gravitational acceleration and $P_{vac}$ the residual pressure above the piston.

In the present work, the piston gauge is a Fluke PG 7601 (piston surface area 980.49 mm$^2$). A turbomolecular pump (Pfeiffer Hipace 300) backed by a dry (oil-free) pump (Pfeiffer ACP 28) provides the vacuum above the piston ($P_{vac}$=0.1 Pa, u($P_{vac}$)=0.1 Pa).

**2.3 Gas compensation system**

The gas compensation system (Fig. 1 (a)), is composed of the gas source, the pressure regulator, ballast volumes, mass flow controllers (MFC), absolute pressure gauge, the laser interferometer, pump together with the requisite pipes and valves.

The gas source is a bottle of high purity (>99.99999%) $^4$He at high pressure (20 MPa). Using the pressure regulator, the gas pressure at the entrance to the 1 L ballast volume is reduced to 1.5 bars (0.15 MPa).

The first mass flow controller MFC1 (S48 300/HMT) is used to fix the mass flow rate transfer to the upstream, usually a small value around 2 sccm[1]. The gas flow is split into two parts using a tee: one part going to supplement the piston gauge leakage while the other is evacuated by a dry pump (Pfeiffer ACP 15). The second mass flow controller MFC2 (S48 300/HMT) is used to control the amount of gas evacuated. Thus, the extra quantity corresponds to the difference in flow difference between the two controllers. A four-way cross is used to connect ballast volume 2 (1 L), the pressure vessel and the piston gauge. In normal operation, since the flow rate from the four-way cross to the piston gauge is very small (about $10^{-2}$ sccm) with almost no gas flowing to the pressure vessel and ballast volume 2, the pressure in the whole system is nearly uniform aside from some hydrostatic head difference. All the test experiments were performed at room temperature (about 24.5 °C), independently of the microwave resonator. The latter was isolated from the system by closing valve 2 and the volume of the pressure vessel simulated using the 1 L ballast volume 2. Upon completion of the test experiments, valve 2 was opened and the pressure control system used to control the pressure within the microwave resonator. To estimate the stability of the controlled pressure, two absolute pressure transducers (Paroscientific Digiquartz 745) of range 0-23 psig with nano- i.e. parts-per-billion ($10^{-9}$) resolution, were used to track the pressure in real time.

---

[1] Sccm=standard cubic centimeters per minute, corresponding to $1.6667 \times 10^{-8}$ m$^3 \cdot$s$^{-1}$ in the International System of Units (SI). We define the volumetric flow 1 sccm as the flow of 1 cubic centimeter per minute of argon at the pressure 103 kPa and temperature 20°C.

During normal operation, MFC1 is used to fix the flow rate at 2 sccm (or some other small value) while the flow rate through MFC2 is determined by the piston height. When the piston height lies above the set value, the set-point of MFC2 is reduced, causing the piston to rise to the chosen height. The reverse happens whenever the piston height lies below the set value. Thus, the piston height remains almost constant with only a very small oscillation. Servo-control using a proportional-derivative (PD) feedback loop is implemented via a control program written in LabVIEW™ software. The piston height is measured by the LVDT height sensor internal to the piston gauge (resolution 10 μm) and an external laser interferometer (Keyence LK-G80, resolution 0.1 μm).

## 3 Experimental results and discussions

### 3.1 Comparisons between rotation and non-rotation behaviour of PG 7601

When the piston gauge is in operation, the piston must float freely without friction due to contact between the piston and cylinder. Thus, it is recommended that, during pressure measurements, the piston should be in permanent rotation. To supply gas pressure for pressure control, the piston must fall. As times goes by, the piston slows down due to viscous damping by the surrounding gas. Whenever the piston reaches its lower rotational speed threshold, the control electronics of the gauge add some angular acceleration, which causes the pressure to jump. In the case of the present experiment, this happens every 30 minutes, which is unacceptable. It was thus decided to operate the gauge without rotation.

To test this approach, the rotational and non-rotational behaviours of the piston-cylinder assembly were compared with the piston falling. The procedure as follows: (1) Close valve 2 and set the flow through MFC2 to 0 sccm. (2) Apply a flow of 2 sccm to MFC1 and wait until the piston reaches its highest point. (3) Set the value of MFC1 to 0 sccm and let the piston fall freely. For the rotation experiment the "auto-rotation function" was enabled with minimum and maximum rotation speeds of 20 rpm and 100 rpm respectively. For the non-rotation experiment, the "auto-rotation function" was disabled.

Fig. 2 shows the results of experiments with and without rotation under a nominal pressure 60 kPa, i.e, for a nominal load mass of 6 kg. Fig. 2 (a) shows the rate of descent of the piston for the two modes of operation where the piston height was measured using the internal height sensor. Both curves are linear and show similar rates of descent (90 µm/min for rotating mode and 97 µm/min for non-rotating mode). Fig. 2 (a) also includes two zooms for different heights. In the zoom on the right, one can see that the curve for the non-rotating piston is smooth while that for the rotating mode exhibits some fluctuations. The reason is that the rotation of the piston is not strictly co-axial so its height varies slightly as it wobbles. This fluctuation interferes with the servo-loop used to stabilize the piston height. On the zoom on the left-hand side, a different zone is visible for the experiment in rotation mode. The reason can be found in Fig. 2 (b): every time the piston is re-accelerated by the driving motor after reaching its minimum permissible rotation speed, the contact between the motor and piston causes the piston to jump upwards.

The contact also upsets the force balance of the piston, thereby causing a significant pressure shift (Fig. 2 (c)). After one minute of acceleration, the piston becomes free once again and a stable pressure is restored. This scenario repeats itself every 1.5 hours. By contrast, the non-rotational mode exhibits no such behaviour. Fig 2 (d) gives comparisons of pressure between the rotating and non-rotating piston experiment. One sees that the pressure increases as the piston falls. The pressure difference between the heights of +2.5 mm and -2.5 mm is 0.75 Pa. According to the calculation of Sharipov *et al.* [27], this difference arises from additional forces on the piston due to the gas flow in the crevice between piston and sleeve. Moreover, the gas pressure beneath the piston and filling the crevices causes an elastic deformation of both sleeve and piston. Though small, it modifies the dimensions, so its effect should be included in the calculation of the effective area.

More importantly, the difference in mean pressure value with and without piston rotation is less than 50 mPa, as is obvious from Fig. 2 (d). This means the choice of non-rotational mode has only a small effect on the measured pressure. Given this and the fact that the non-rotational mode is immune to periodic jumps in piston height that hinder servo-control, this mode was adopted in the remainder of the experiments described here.

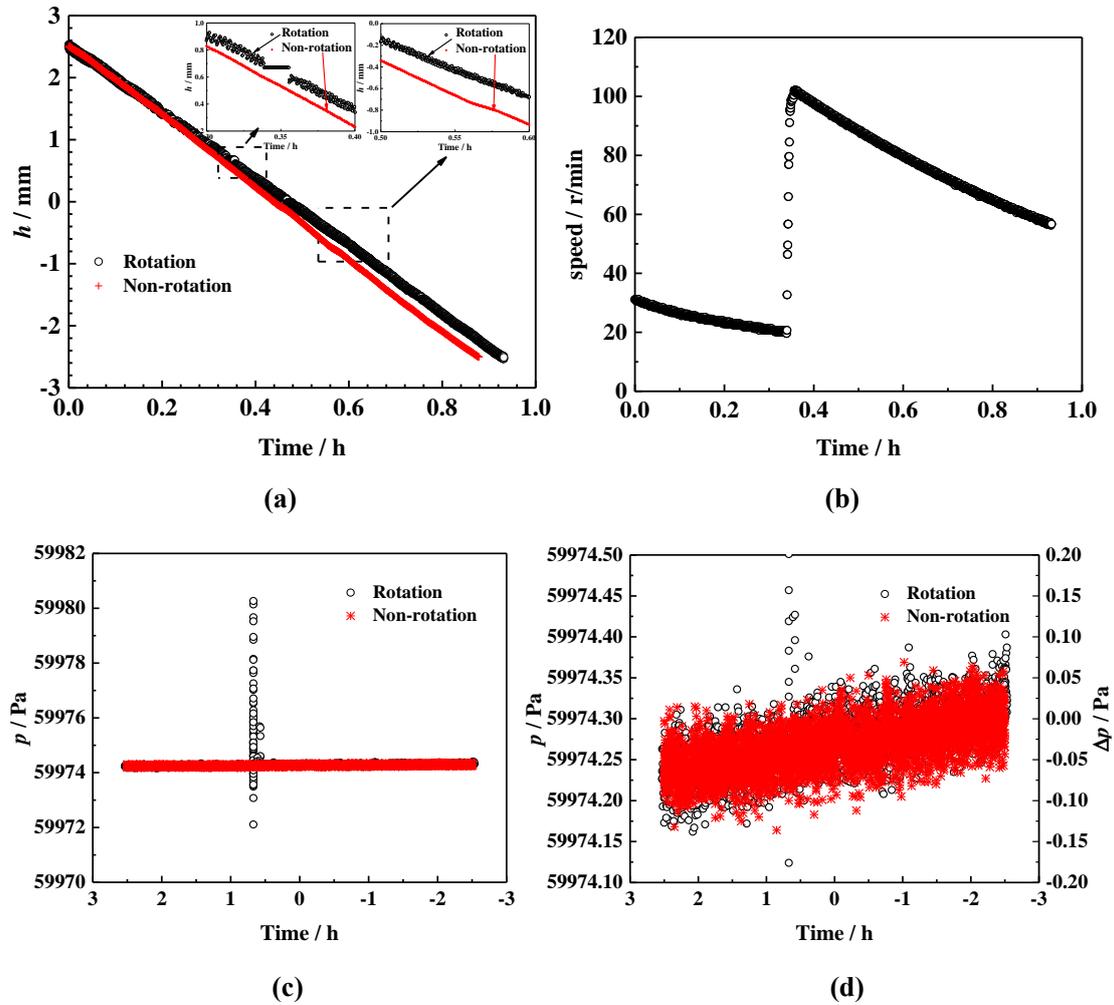

**Fig. 2** Descent of the piston in rotation and non-rotation experiments (a) Height profile (b) Speed profile (c) Pressure profile of rotation experiment (d) Comparisons of pressure between rotation and non-rotation experiment ("Each point corresponds to a 44 ms integration time.")

### 3.2 Comparison of falling and regulatory behaviour of the piston cylinder gauge

It is generally recommended that the piston gauge be operated at a small, reproducible falling rate. In the present application, however, the fall of the piston causes two problems. First, after dropping for one or two hours, the piston reaches its lowest height below which it can no longer control the pressure. Second, the pressure changes by $10^{-1}$ Pa (Fig. 2 (d)) as the piston falls, which makes it generate an additional pressure instability. To solve these problems, the gas compensation system mentioned in section 2.2 was installed to regulate the piston height.

To compare the falling and regulation modes of operation of the piston gauge. The piston was regulated in non-rotation mode at different heights between +2.5 mm and -2.5 mm in steps of 0.5 mm. Fig. 3 shows the results of the regulation and falling piston experiments.

As shown in Fig. 3 (a), for regulation mode, the generated pressure increases as the height decreases. The behaviour follows the same trend as for the falling piston experiment. However, the regulated pressure is always a little larger than that measured in the latter. To quantify the pressure difference, we fitted the pressure measured in falling mode by a second-order polynomial and compared the pressure values for the different regulated heights. It is apparent from Fig. 3 (b) that the pressure difference between regulation and rotation-falling modes is less than 0.1 Pa. This difference is very small, almost the same as that between h =2.5 mm and h = -2.5mm in falling mode.

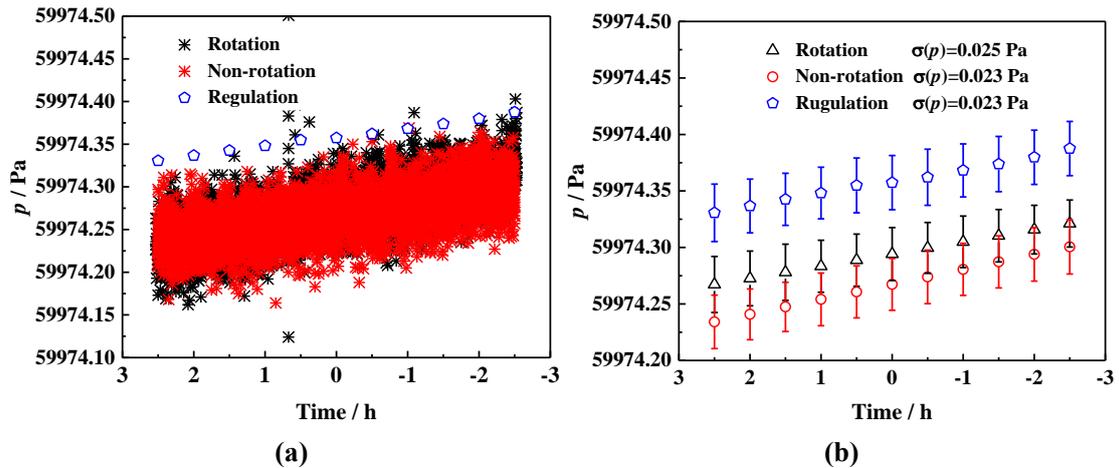

**Fig. 3 Piston falling and regulation experiments (a) Original pressure profile (b) mean pressure values for different piston heights (The regulation is performed in at non-rotation mode using the internal LVDT height sensor. All pressure values are measured by the absolute pressure transducer with an integration time 44 ms )**

From the above discussions, it is found that, for the piston gauge, the "regulation non-rotation mode" yields a pressure that differs from that measured in the "falling rotation mode" by less than 0.1 Pa.

Since the "regulation non-rotation mode" avoids the problems caused by piston fall and rotation stated above, and provides a stable long-term pressure, the remaining experiments were all carried out using it. For completeness, the corresponding uncertainty arising from the small pressure difference between the two modes is included in the uncertainty budget (section 4.1).

**3.3 Pressure regulation at room-temperature using a ballast volume**

The main goal of the pressure control system was to provide a stable pressure for SPRIGT, i.e. primary thermometry at cryogenic temperatures (4 K-24.5 K). Indeed, this was achieved, as described later. In a first instance, however, pilot experiments were performed at room temperature and some improvements made as a result of them.

For the room temperature studies, the pressure vessel containing the microwave resonator in the cryostat was isolated by closing valve 2 (Fig. 1 (a)). The setpoint of mass flow controller MFC1 was chosen to be 2 sccm set point while that of MFC2 was determined in real time by feedback control software written in LabVIEW™. The purpose of the feedback loop is to stabilize the piston at a constant height. In this experiment, the feedback signal is derived from the height measured by the internal sensor of the piston gauge with a resolution of 10 μm.

Fig. 4 shows the results. In Fig. 4 (a), the piston height appears very steady with small jumps (0.01 mm). In reality, the piston rises or falls very smoothly. Since the resolution of LVDT is 10 μm, its reading number is deceptive. Because, via the PID program, the set-point of MFC 2 is determined by the piston height signal with small jumps, sudden changes can occur in the flow through MFC2 (see Fig. 4 (b)). Fig. 4 (c) shows that the pressure, measured by the absolute pressure gauge with an integration time of 44 ms, fluctuates with a relative standard deviation of 0.4 ppm. To measure the real noise, namely the instability, the white noise should be removed. This is easily achieved by increasing the integration time of the absolute pressure gauge. Figs. 4 (d), (e) and (f) show the pressure measured with integration times of 178 ms, 700 ms and 1400 ms. The noise decreases as the integration time is increased, but the noise changes rarely between results for integration times of 700 ms and 1400 ms. This means the noise measured with an integration time 700 ms corresponds to the real noise.

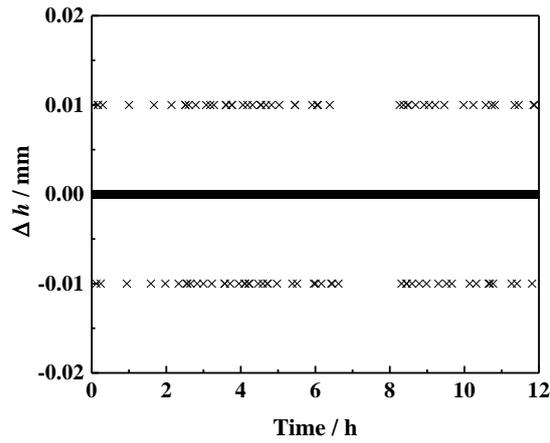
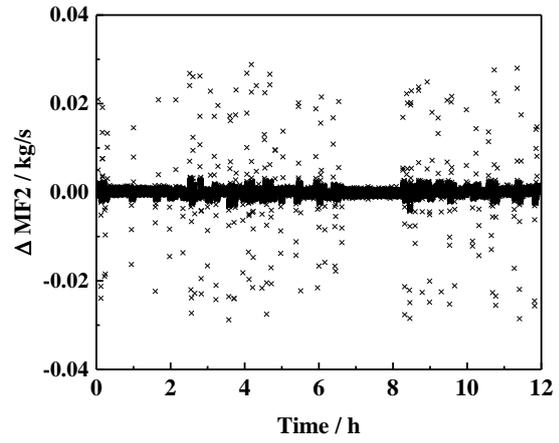

(a)                                                  (b)

(c)                                                  (d)

(e)                                                  (f)

**Fig. 4 Piston height control results based on piston heights measured by the internal sensor (a) Piston height (b) Flow of mass flow controller 2 (c) 44 ms (d) 178 ms (e) 700 ms (f) 1400 ms ((c), (d), (e) and (f) are the controlled pressures measured by the absolute pressure transducer (Digiquartz1) at 60 kPa for the different integration times)**

One can see in Fig. 4 that some points lie far from the mean, no matter how long the integration time. The excursions are caused by the sudden flow changes due to the aforementioned low resolution (10 μm) of the LVDT height sensor. This means a better resolution is required for piston height control.

To measure piston height more precisely, a laser interferometer (Keyence LK-G80, resolution 100 nm) was installed above piston gauge. Specifically, a mirror the upper surface of the piston forms part of the optical path. Fig. 5 shows the pressure control results obtained using it. The piston height is stabilized to within 4 μm of the set point, a clear improvement upon the use of the internal LVDT sensor. As a result, the pressure is controlled with instability of 0.075 ppm, clearly better than that shown in Fig .4 (b).

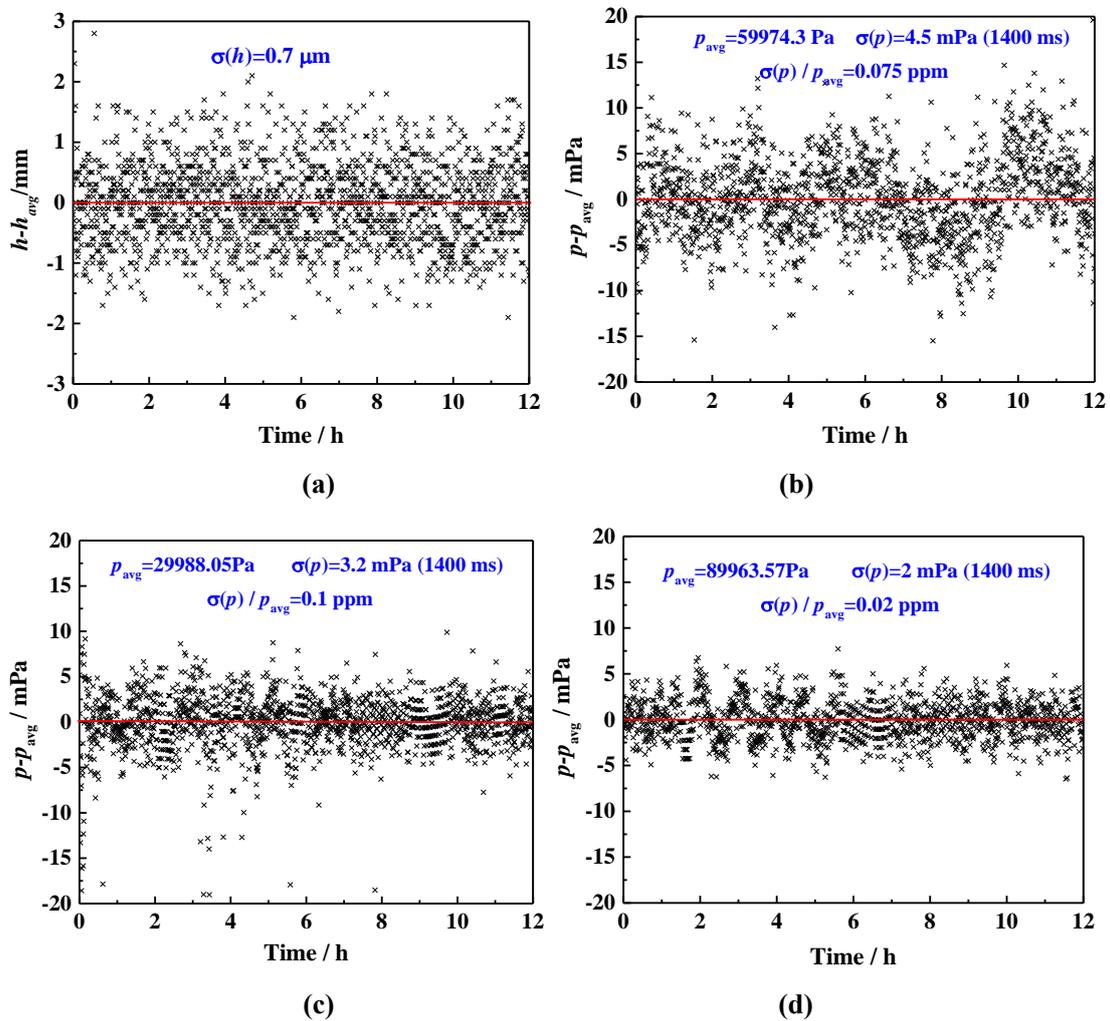

**Fig. 5 Piston height and pressure control results based on piston heights measured by external interferometer (a) Piston height (b) Controlled pressure near 60 kPa (c) *Idem* near 30 kPa (d) *Idem* near 90 kPa**

### 3.4 Low-temperature regulation of the microwave resonator gas pressure

Following the test experiments at room temperature, the pressure regulation system was connected to the resonator in the cryostat to provide a stable pressure at low temperature. The resonator was connected by opening valve 2 shown in Fig. 1(a). As with the experiments at room temperature, the flow through the mass-flow controller MFC1 is fixed while that through MFC2 is determined in real time by the feedback signal derived from the height of the piston gauge. Before the pressure control system is connected to the cryostat, the pressure system should be initially under vacuum, after which the gas is slowly introduced into the pressure system and the resonator in the cryostat. As shown in Fig. 6, the relative fluctuations at pressures near 30 kPa, 60 kPa 90 kPa and 120 kPa in the temperature range 5 K-24.5561 K are all less than 0.16 ppm, i.e. 25 times lower than the 4 ppm pressure stability required for SPRIGT at the 0.25 mK level of uncertainty.

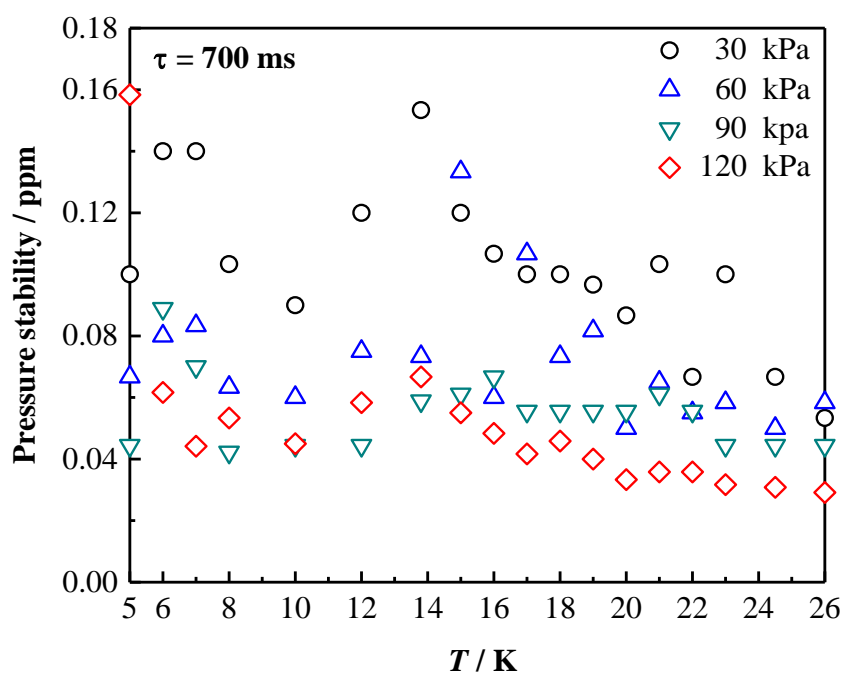

Fig. 6 Pressure stability at different pressures for temperatures from 5 - 25 K

## 4 Uncertainty Budget

### 4.1 Room Temperature pressure regulation

The values and uncertainties calibrated by the National Institute of Metrology of China used in the determination of pressure $p$ are listed in Table 1. The area of the piston gauge was calibrated at TIPC by NIM in 2019 at 24.5 °C. A correction for thermal expansion of piston and cylinder calibrated by NIM is applied to calculate the $A_{\theta, p}$.

Table 1 Uncertainty budget for pressure using the piston gauge determinations (k=1)

|  |  | Mean | Standard deviation |
|---|---|---|---|
| Piston mass/kg | | 0.40000109 | 6E-07 |
| Carrying bell mass/kg | | 0.2999747 | 5E-07 |
| Weights mass/kg | (30 kPa) | 2.3000355 | 3.02E-06 |
| | (60 kPa) | 5.3000425 | 8.01E-06 |
| | (90 kPa) | 8.3000895 | 8.70E-06 |
| | (120 kPa) | 11.300108 | 1.14E-06 |
| $g$/ m·s$^{-2}$ | | 9.80111294 | 9.80E-07 |
| $A_{24.5,0}$/mm$^2$ | | 980.5216 | 5.00E-03 |
| $\theta$ / °C | | 24.7 | 0.01 |
| $\alpha_p$ /°C$^{-1}$ | | 3.6E-06 | 1E-06 |
| $\alpha_c$ /°C$^{-1}$ | | 3.6E-06 | 1E-06 |
| $P_{vac}$/Pa | | 0.1 | 0.1 |
| $\Delta P_{regulation}$ / Pa | | | 0.06 |

The uncertainty budget for the pressure measurement of the piston gauge $P_A$ at room temperature is shown in Table 2. Within the uncertainty caused by the regulation experiment with different heights $P_{regulate}$, the $P_A$ combined standard uncertainties are large at low pressures.

The combined standard uncertainty for $P_A$ is obtained by quadratic summing of the estimated uncertainty components listed in Table 2. Under the present absolute mode of operation of the piston gauge, the pressure reproducibility is not only related to the masses of load and piston, gravity and piston effective area, but also affected by the vacuum in the region above the pressure balance, the thermal expansion coefficients of piston and cylinder and their temperature. Once the pressure control system is connected to the cryostat, the uncertainty in the hydrostatic head correction $P_{HA}$ must also be included. In the uncertainty budget for pressure measurements at room temperature, this sub-component will be neglected. Eq. (1) and Eq. (2) give the mathematical model used to deduce the pressure using the PG7601 piston gauge, taking into account the aforementioned effects. The effective area of the piston is given by

$$A_{\theta,P} = A_{24.5,0} \cdot 10^6 \cdot [1+(\theta-24.5)(\alpha_P+\alpha_C)](1+\lambda P_{nom}) \qquad (2)$$

where $A_{24.5,0}$ (in mm$^2$) is the effective piston area at 24.5 °C and zero pressure; $\theta$ the piston temperature, °C; $\alpha_p$ and $\alpha_c$ the respective thermal expansion coefficients of piston and cylinder, °C$^{-1}$; $\lambda$ the elastic deformation coefficient of the piston-cylinder assembly, MPa$^{-1}$; and $P_{nom}$ is the nominal pressure, MPa.

The dominant uncertainty component at all pressures measured in the present study arises due to the 5 ppm (k=1) uncertainty in the effective area of the balance piston. The uncertainty in the regulation experiment for different piston heights is estimated from the descending rotating piston experiments. A drop in piston height from +2.5 mm to -2.5 mm corresponds to a pressure difference of less than 0.1 Pa. Therefore, the different height regulation correction corresponds to a 0.06 Pa uncertainty. The value of the local gravitational acceleration $g$ and its uncertainty were measured and calculated by the National Institute of Metrology of China.

**Table 2 Uncertainty budget for the pressure of piston gauge determinations (k=1)**

|  | u(x) / Pa | | | |
|---|---|---|---|---|
|  | P=30 kPa | P=60 kPa | P=90 kPa | P=120 kPa |
| Mass_piston / kg | 0.006 | 0.006 | 0.006 | 0.006 |
| Mass_carryingbell / kg | 0.005 | 0.005 | 0.005 | 0.005 |
| Mass_weights / kg | 0.030 | 0.08 | 0.087 | 0.11 |
| $g$ / m·s$^{-2}$ | 0.003 | 0.006 | 0.009 | 0.012 |
| $A_{24.5,0}$/mm$^2$ | 0.15 | 0.30 | 0.45 | 0.6 |
| $\theta$/°C | 0.002 | 0.004 | 0.006 | 0.009 |
| $\alpha_p$ /°C$^{-1}$ | 0.006 | 0.012 | 0.018 | 0.024 |
| $\alpha_c$ /°C$^{-1}$ | 0.006 | 0.012 | 0.018 | 0.024 |
| $P_{vac}$/Pa | 0.1 | 0.1 | 0.1 | 0.1 |
| $\Delta P_{regulation}$/Pa | 0.06 | 0.06 | 0.06 | 0.06 |
| u$P_A$ / Pa | 0.20 | 0.33 | 0.47 | 0.62 |

### 4.2 Pressure regulation of the resonator at cryogenic temperature

When the pressure control system is connected to the resonator in the cryostat, one must include the uncertainty in the thermodynamic temperature due to the pressure measuement. This contains two contributions: the absolute value of the pressure and the noise due to imperfect regulation.

The uncertainty due to the absolute pressure is calculated by quadrature summation of two sub-components assumed to be statistically independent. The first is due to the calibration of the piston gauge which is the same as the calculation for room temperature measurements. The second sub-component is the uncertainty in the hydrostatic head correction from the gas line. For the present study, the hydrostatic head corrections are calculated using Eq. (3). The gas line was divided into $n$ straight segments. The thermal gradients along the gas line inside the cryostat between these measured endpoint temperatures are estimated using a simple linear approximation. The uncertainty in the hydrostatic head correction contribution from the pressure tube dominated by the uncertainty in the height change (calculated from the uncertainties in the length measurements), and multiplied with the average gas density in the segment. Thus, the segments that contribute most strongly to the overall hydrostatic head correction uncertainty are those near the low-temperature end of the gas line that has large height-change uncertainties. The uncertainty due to imperfect pressure regulation varies with temperature, and is calculated using the standard deviation of the pressure regulation results:

$$P_{HA} = -g \sum_i (\rho_i \Delta h_i) \tag{3}$$

where $\rho_i$ is the average gas density in segment $i$; $\Delta h_i$ is the change in height of the $i^{th}$ segment.

As shown in Table 3, the dominant uncertainty component due to pressure for SPRIGT thermodynamic temperature measurement at all temperatures arises from the uncertainty in the pressure calibration of the piston gauge and the pressure stability.

Table 3 Uncertainty budget of P for SPRIGT T determinations ($P$ = 60 kPa, k=1)

| T / K | $P_A$ /mK Calibration | $P_{HA}$ /mK Hydrostatic | $p$ stability /mK | $p_{total}$ /mK |
|---|---|---|---|---|
| 26 | 0.008 | 0.00053 | 0.0015 | 0.008 |
| 24.5561 (Ne triple point) | 0 | 0 | 0.0025 | 0.002 |
| 23 | 0.008 | 0.00057 | 0.0013 | 0.008 |
| 22 | 0.012 | 0.00092 | 0.0012 | 0.012 |
| 21 | 0.016 | 0.00129 | 0.0014 | 0.016 |
| 20.26 | 0.019 | 0.00154 | 0.0011 | 0.019 |
| 19 | 0.023 | 0.00201 | 0.0016 | 0.023 |
| 18 | 0.026 | 0.00235 | 0.0013 | 0.026 |
| 17 | 0.028 | 0.00271 | 0.0018 | 0.028 |
| 16 | 0.029 | 0.00306 | 0.0010 | 0.030 |
| 15 | 0.031 | 0.00343 | 0.0020 | 0.031 |
| 13.8 ($H_2$ fixed point) | 0.032 | 0.00384 | 0.0010 | 0.032 |
| 12 | 0.032 | 0.00448 | 0.0008 | 0.032 |
| 10 | 0.030 | 0.00520 | 0.0006 | 0.030 |
| 8 | 0.026 | 0.00595 | 0.0005 | 0.027 |
| 7 | 0.024 | 0.00634 | 0.0006 | 0.025 |
| 6 | 0.021 | 0.00678 | 0.0005 | 0.022 |
| 5 | 0.018 | 0.00730 | 0.0003 | 0.019 |

## 5 Conclusion

A high stability pressure regulation system for helium-4 gas based on a micro-flow compensation system has been developed and applied at both room temperature and cryogenic temperatures (5-25 K). Its purpose is to allow single-pressure refractive index gas thermometry in the latter range at the 0.25 mK level of accuracy. Since the refractive index is a function of both temperature and pressure, the latter must be controlled to better than 4 ppm. The main conclusions are as follows:

1. The pressure control system incorporates a piston gauge (Fluke PG 7601) whose operation in both rotation (falling) mode and non-rotating (regulation) mode was investigated in this paper. It was found that the mean pressure difference is less than 0.05 Pa, and that the non-rotation mode avoids large pressure jumps caused by the speeding up of piston rotation whenever it reaches the low revolution threshold.

2. Active servo control to maintain the piston gauge at constant height was studied using both the internal height sensor of the gauge and an external laser interferometer. The results for room temperature showed that the pressure instability, defined by standard deviation, was of order 3 mPa at 30 kPa, 4.5 mPa at 60 kPa and 2 mPa at 90 kPa, while in the range 5 K-24.5561 K, it is better than 0.16 ppm.

3. The uncertainty budget for the pressure of piston gauge determinations and the

uncertainty contribution to SPRIGT were all analyzed. The dominant component (less than 0.032 mK) is that due to the measurement of absolute pressure.

**ACKNOWLEDGEMENTS**

This work was supported by the National Key Research and Development Program of China (grant n° 2016YFE0204200), the National Natural Science Foundation of China (grant n° 51627809), the International Partnership Program of the Chinese Academy of Sciences (grant n° 1A1111KYSB20160017) and the European Metrology Research Program (EMRP) Joint Research Project 18SIB02 "Real K". The authors are also grateful for the expert technical assistance of Zhen Zhang and Ying Ma.

**NOMENCLATURE**

| | |
|---|---|
| $A_{20,0}$ | Piston effective area at 20 °C and zero pressure [mm$^2$] |
| $A_{\theta,p}$ | Piston effective area [m$^2$] |
| $g$ | Local gravitational acceleration [m·s$^{-2}$] |
| $H$ | Piston height indicated by the laser interferometer [mm] |
| $h$ | Piston height [mm] |
| $M$ | Mass [kg] |
| MF2 | Mass flow [kg·s$^{-1}$] |
| $P_A$ | Absolute pressure of piston gauge (below piston) [Pa] |
| $P_{HA}$ | Fluid head correction [Pa] |
| $P_{nom}$ | Nominal pressure [Pa] |
| $\Delta P_{regulation}$ | Pressure difference between regulation and falling piston modes [Pa] |
| $P_{vac}$ | Residual vacuum pressure above the piston [Pa] |

Greek characters

| | |
|---|---|
| $\theta$ | Piston temperature [°C] |
| $\alpha_p$ | Thermal expansion coefficient of piston, [°C$^{-1}$] |
| $\alpha_c$ | Thermal expansion coefficient of cylinder, [°C$^{-1}$] |
| $\lambda$ | Elastic deformation coefficient of the piston-cylinder, [1·MPa$^{-1}$] |
| $\rho_i$ | Average gas density in segment $i$, [kg·m$^{-3}$] |
| $\Delta h_i$ | Change in height of segment $i$ [m] |